\documentclass[twocolumn,showpacs,english,superscriptaddress,preprintnumbers,amsmath,amssymb,floatfix,groupedaddress]{revtex4-1}

\usepackage[T1]{fontenc}
\usepackage[latin9]{inputenc}
\usepackage{dcolumn}
\usepackage{bm}
\usepackage{graphicx}
\usepackage{color}
\usepackage{esint}
\usepackage{babel}
\usepackage{amsfonts}
\usepackage{slashed}
\usepackage{enumerate}

\newcommand{\beq}{\begin{eqnarray}}
\newcommand{\eeq}{\end{eqnarray}}
\newcommand{\beqq}{\begin{eqnarray*}}
\newcommand{\eeqq}{\end{eqnarray*}}

\usepackage{graphicx}
\usepackage{dcolumn}
\usepackage{bm}
\usepackage{color}
\usepackage{tabularx}

\usepackage{amsmath}

\begin{document}

\begin{titlepage}

\title{Two-Channel Kondo Physics in a Majorana Island Coupled to a Josephson Junction}
\author{L. A. Landau}
\author{E. Sela}
\affiliation{Department of Physics and Astronomy, Tel Aviv University}

\begin{abstract}
We study a Majorana island coupled to a bulk superconductor via a Josephson junction and to multiple external normal leads. In the absence of the Josephson coupling, the system displays a topological Kondo state, which had been largely studied recently. However, we find that this state is unstable even to small Josephson coupling, which instead leads at low temperature $T$ to a new fixed point. Most interesting is the case of three external leads, forming a minimal electronic realization of the long sought two-channel Kondo effect.
%We map out the phase diagram and compute the low $T$ conductance.
While the $T=0$ conductance corresponds to simple resonant Andreev reflection, the leading $T$ dependence forms an experimental fingerprint for non-Fermi liquid properties.
\end{abstract}


\maketitle

\draft

\vspace{2mm}

\end{titlepage}

\section{Introduction} \label{Introduction} Majorana fermions are zero energy spatially localized states that emerge in topological superconductors as an equal superposition of electrons and holes \cite{kitaev,read,wilczek}. Theoretical predictions for their presence in nanoscale devices, such as spin-orbit coupled wires in proximity to a superconductor  \cite{sarma,oreg,weizmann}, were strongly supported by recent experiments  \cite{kouwenhoven, heiblum, marcus}. One of the significant properties of a superconducting island hosting Majorana modes is its ground state degeneracy. However, when such an island is in the Coulomb blockade regime this degeneracy is lowered and replaced by two sectors of distinct charge parity, each of which has either an even or an odd number of electrons \cite{fu,xu}.

Recent theoretical progress \cite{ BeriCooper,AltlandEgger,Beri,Affleck13,AltlandBeryEggerTsvelik14,Tsvelik,numerics,ZazunovAltlandEgger14,Eriksson14a,Eriksson14,  Kashuba15,Pikulin16,Meidan16,Plugge16,mora,michaeli,beek,vanHeck}, particularly the works of B{\'e}ri-Cooper \cite{BeriCooper} and Altland-Egger \cite{AltlandEgger} have paved the way for the study of such Majorana islands, predicting the emergence of a ``topological Kondo effect'' in the Coulomb valley regime. Under the  condition where the number of electrons in the island is fixed, and where the number of lead-coupled Majorana modes exceeds two, $M>2$, the Majorana degrees of freedom non-locally encode an effective quantum impurity spin. This ``spin'' collectively interacts with the lead's electrons, leading to a correlated state characterized by non-Fermi liquid (NFL) behavior that is observable in the electrical conductance. Recently, it was shown that this behavior emerges at much higher temperatures near charge degeneracy points \cite{michaeli,mora}. Motivated by this and following Ref.~\onlinecite{Eriksson14}, we ask the question: what are the consequences of breaking charge conservation on the properties of the system?

\begin{figure}[pt]
	\centering
	\includegraphics[scale=0.3]{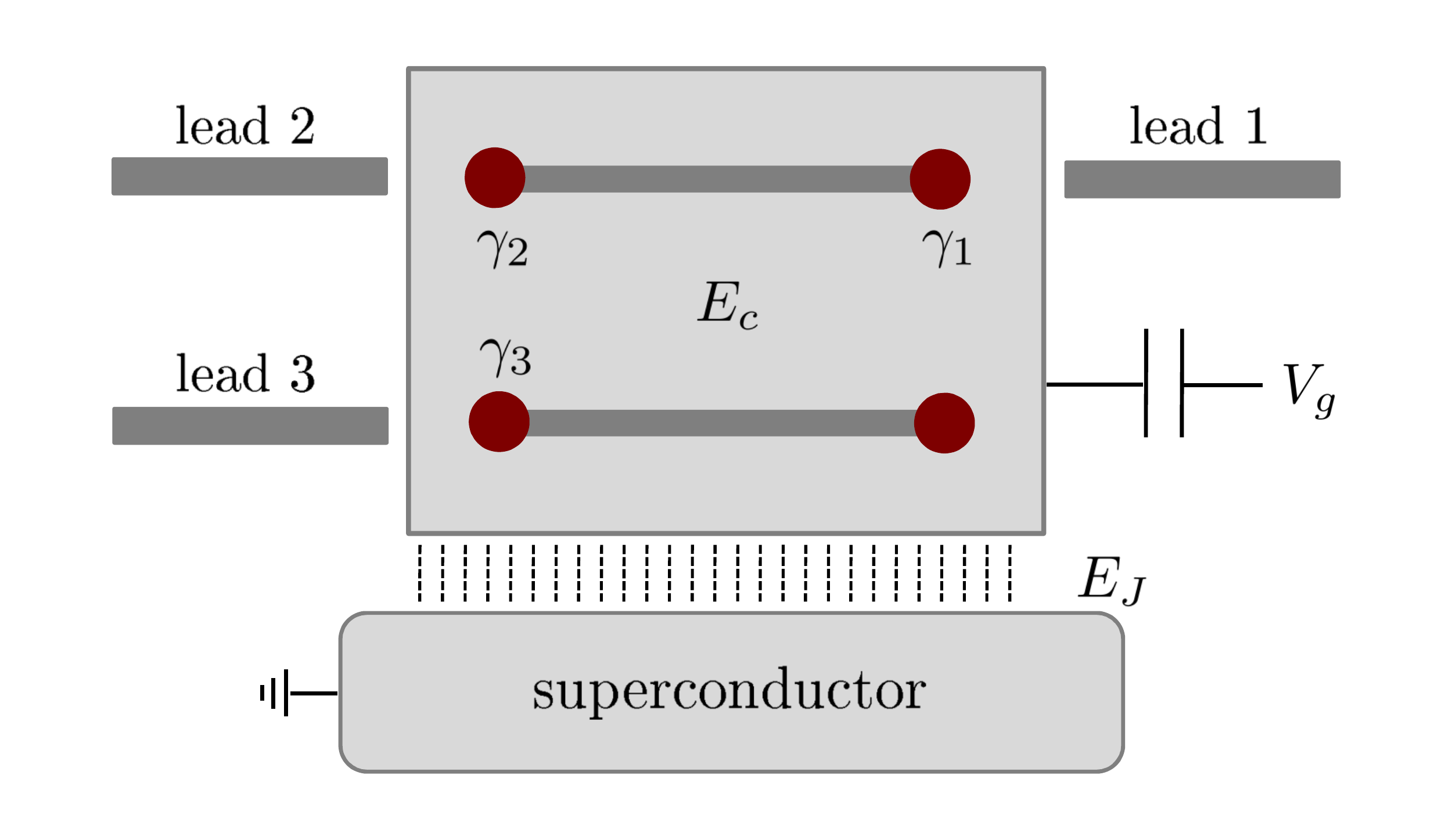}
		\caption{Schematics of the device: superconducting island with spatially localized Majorana modes and with charging energy $E_c$, coupled to a bulk superconductor and to $M$ (here $M=3$) external normal leads. A gate voltage tunes the number of electrons in the island.    }
\label{fig1}
\end{figure}

Our work deals with multi-terminal charge transport through a Majorana island connected to $M>2$ external leads via Majorana tunneling junctions. In addition, the island is coupled via a Josephson junction to a grounded bulk superconductor (SC), see Fig.~\ref{fig1}. Having sufficiently long wires, we assume that the Majoranas have no direct coupling. While the charge in the island is tuned by a gate voltage, the Josephson coupling allows charge fluctuations in units of $2e$ between the island and the bulk superconductor. 

The aforementioned topological Kondo state is known to be completely stable against lead asymmetry~\cite{BeriCooper} and gate voltage detuning~\cite{michaeli,mora}. In this paper, however, we show that the Josephson coupling gives rise to an instability of the topological Kondo fixed point. In a charge conserving system and far from a charge degeneracy point, tunneling events between the leads and the island are of the form $\psi_i^\dagger \psi_j$ which merely describes the exchange of charge $e$ between leads $i$ and $j$. Here $\psi_i$ ($\psi_i^\dagger$) annihilates (creates) an electron in lead $i$. However, the lack of charge conservation permits tunneling events of $\it{two}$ electrons from the leads to the island and then to the bulk SC, leading to anomalous terms of the form $\psi_i \psi_j$ (or $\psi_i^\dagger \psi_j^\dagger$). As our analysis shows, these terms can be identified with channel anisotropy in the topological Kondo Hamiltonian and as a result the system is driven towards a new fixed point of strong coupling regime. The equal combination of these tunneling events which effective emerges at low temperature, leads to a correlated Kondo state involving only one Majorana field $\psi_i\pm\psi_i^\dagger$ from each lead. For $M=3$ for example, this is equivalent \cite{Tsvelik1, Tsvelik2} to two-channel Kondo (2CK) physics allowing for new ways to explore its non-Fermi liquid properties.

%At the new fixed point, we find that the low energy physics of the island is dominated by $\it{parity}$ interaction. Different than charging energy, the parity interaction distinguishes states just according to their parity. However, as we show its magnitude can be of the order of the charging energy.  The parity operator can be expressed as the product of all Majorana operators $i^{N/2}\prod_{j}{\gamma_j}$, leading to a non-local correlation between tunneling events at different leads.
The full phase diagram of the system can be mapped as function of the ratio of the Josephson energy $E_J$ to the charging energy $E_c$ and as function of temperature $T$.  As we show, at $E_J \neq 0$, where the system flows to a new fixed point, the zero-temperature conductance is $\frac{2e^2}{h}$ and associated with Andreev reflection. Furthermore, low temperature corrections to the conductance have a universal power-law dependence $T^\alpha$ with  $\alpha=1$ for $M=3$. This provides an experimental signature of the non-Fermi liquid behavior of the 2CK state. On the other hand, we find $\alpha=2$ for all $M>3$. 

The rest of the paper is organized as follows: in Sec.~\ref{Model} we present a detailed formulation of the model. Sec.~\ref{Phase diagram} presents the emergent instability at the Kondo fixed point and parity interaction, and includes the phase diagram of the system. In Sec.~\ref{Low energy conductance} we deal with calculation of the low energy conductance, its low $T$ corrections and the effect  of interactions in the leads. We summarize in Sec.~\ref{Summary}
% we consider the properties of the new fixed point, and show the different regimes where parity interaction takes place.

\section{Model}  \label{Model} Our system is described by the Hamiltonian $H =  H_{\rm{c}}+H_{\rm{J}}+ H_{\rm{leads}} +H_{\rm{T}}$. %The leads are assumed to be spinless with density of states $\rho$ at the Fermi level.
The island is coupled to a gate enforcing both its charging energy $E_c$ and average occupancy, leading to the charging Hamiltonian
\begin{align}
H_{\rm{c}} = E_c \left(\mathcal{N}-n_g\right)^2,
\end{align}
where $\mathcal{N}$ is the electron number operator relative to the gate voltage parameter $n_g$. In conventional superconductors, the ground state is expected to have an even number of electrons due to the superconducting energy gap required by an unpaired electron. However, the hosted zero energy Majorana modes indeed allow odd occupancy without paying this energy. By tuning the gate voltage the number of electrons in the ground state is fixed to be the integer which is closest to $n_g$, denoted by $N_0$. However, when $n_g$ has half integer values the ground state is degenerated and consists of two states whose charges differ by $e$.\\
The Josephson coupling between the island and the bulk superconductor enables Cooper-pair tunneling described by
\begin{align}
H_{\rm{J}} = -E_J\cos(\phi), \label{HJ}
\end{align}
where $E_J$ is the Josephson energy. The superconducting phase of the island $\phi$ is canonically conjugate to its electron number $\mathcal{N}$ and satisfies the commutation relation $[\phi,\mathcal{N}]=2i$. Tunneling of a Cooper-pair into the island conserves its parity but changes its charge by $\pm 2e$.\\
In addition, single-electron tunneling between the island and external leads is described by
\begin{align}
H_{\rm{T}} =\sum_{j=1}^M t_j \psi_j^\dagger(0) \gamma_j e^{-i \phi/2}+{\rm{h.c.}}, \label{HT}
\end{align}
where $\psi^\dagger_j(0)$ is a creation operator of a single electron at the endpoint of lead $j$. The neutral Majorana operators $\gamma_j$'s anti-commute and satisfy $\gamma_j^2=1$. We assume that the Majorana zero modes are far apart and have no direct coupling as the topological quantum wires are sufficiently long. Here, $e^{\pm i \phi/2}$ changes the charge of the island by $\pm e $, i.e half of a Cooper pair, and by contrast to the Josephson coupling, flips its parity. \\
The lead's electrons are modeled by a one dimensional Hamiltonian of non-interacting chiral fermions
\begin{align}
H_{\rm{leads}} =-i\sum_{j=1}^M \int_{-\infty}^{\infty}{\frac{\rm{dx}}{2\pi} v_F \psi_j^\dagger \partial_x \psi_j}, \label{Hleads}
\end{align}
where $\{\psi_i(x),\psi^\dagger_j(x') \}=i\delta_{ij}\delta(x-x') $.

When $E_J=0$, the model reduces to the extensively studied topological Kondo model. Below, we will study the effect of the Josephson term on the properties of the system.
Furthermore, two distinct situations emerge depending on the gate voltage: when $n_g \approx \frac{1}{2}+{\rm{integer}}$, the system is close to a charge degeneracy point and we refer to this as an ``on-resonance" situation, in contrast to the off-resonance case which occurs otherwise. %We shall briefly comment on the effect of interactions later on.

\section{Phase diagram} \label{Phase diagram} In this chapter we map out the phase diagram as function of $T$ and the ratio $E_J/E_c$. This will allow us to connect the charging dominated regime $E_c \gg E_J$, where most previous work has been done, with the Josephson-dominated regime, $E_J \gg E_c$.

\subsection{charging dominated regime}

As a first step, in this subsection we consider the effect of the Josephson coupling as a perturbation and analyze the stability of the Kondo fixed point of the system for $E_J\ll E_c$. To keep the presentation simple, consider a gate voltage very close to the off-resonance point, $n_g \approx \rm{integer}$ (away from charge degeneracy points), and also equal lead's couplings, $t_1=t_2=\ldots=t_M=t$. We apply perturbation theory around zero Josephson coupling, where the unperturbed ground state has a fixed number of electrons $N_0$ (Note that for $M>2$ this is not a unique state, since there are $2^{M/2-1}$ states with fixed charge $N_0$). However, since $H_J$ does not conserve charge, the true ground state cannot have a fixed number of electrons and instead, it consists of superposition of different charge states
\begin{align}
|gs\rangle\approx|N_0\rangle+\frac{E_J}{4E_c}(|N_0+2\rangle+|N_0-2\rangle)+\mathcal{O} \left( \frac{E_J^2}{E_c^2 } \right).
\end{align}
The parity sector subspace including the two states with $ N_0\pm 1 $ electrons in the island is described by a $2\times 2$ Hamiltonian
\begin{align}
H = \left(
      \begin{array}{cc}
        E_c & -E_J \\
        -E_J & E_c  \\
      \end{array}
    \right).
\end{align}
Hence, the two excited states $|ex+\rangle$ and $|ex-\rangle$ include two charge states
\begin{align}
|ex\pm\rangle\approx\frac{1}{\sqrt{2}}(|N_0+1\rangle\pm|N_0-1\rangle)+\mathcal{O} \left( \frac{E_J}{E_c } \right),
\end{align}
with energy eigenstates $E_c\mp E_J$. For weak lead-island coupling $\Gamma=2\pi t^2\nu\ll E_c$, where $\nu$ is the density of states in the leads, the low energy physics of the system is governed by virtual transitions from the ground state to higher charge states.

 Using a Schrieffer-Wolff transformation, we perform leading order perturbation theory in the leads coupling $t$ to obtain an effective Hamiltonian $H_{\rm{eff}}=\langle gs|H_{\rm{T}}\frac{1}{E-H_c-H_J}H_{\rm{T}}|gs\rangle$.  Importantly, the Josephson coupling allows virtual transitions  $|N_0+1\rangle\leftrightarrow|N_0-1\rangle$ or $|N_0\rangle\leftrightarrow|N_0\pm 2\rangle$. This gives rise to terms of the form $\sim\psi_{i} \psi_{j}$ (or $\sim\psi_{i}^{\dagger}\psi_{j}^{\dagger}$), where $\it{two}$  electrons tunnel into (or out of) the island. Taking this into account, the resulting low energy effective Hamiltonian is
\begin{align}
\label{HEFF0}
H_{\rm{eff}}=\frac{t^2}{E_{c}}\sum_{i\neq j}\gamma_{j}\gamma_{i}\left[\psi_{i}^{\dagger}\psi_{j}-\frac{3E_{J}}{2E_{c}}\psi_{i}^{\dagger}\psi_{j}^{\dagger}\right]+\rm{h.c}.
\end{align}\
Expressing the fermionic lead operators as $\psi_j(x)=\frac{1}{\sqrt{2}}(\eta_j(x)+i\rho_j(x))$, where $\rho_j(x)=\rho^\dagger_j(x)$ and $\eta_j(x)=\eta^\dagger_j(x)$ are Majorana operators, $H_{\rm{eff}}$ takes the form
\begin{align}
\label{HEFF1}
H_{\rm{eff}}=J_{\eta}\sum_{i\neq j}\gamma_{j}\gamma_{i}\eta_{i}(0)\eta_{j}(0) + J_{\rho}\sum_{i\neq j}\gamma_{j}\gamma_{i}\rho_{i}(0)\rho_{j}(0),
\end{align}
where $J_{\rho/\eta}=\frac{t^2}{E_c}(1\pm\frac{3E_{J}}{2E_{c}})$. One can see that each Majorana sector $\eta$ and $\rho$  in the leads provides a separate screening channel operator $\eta_{i}(0)\eta_{j}(0)$ or $\rho_{i}(0)\rho_{j}(0)$ coupled to the impurity degree of freedom $\gamma_{j}\gamma_{i}$. Each of these operators satisfies separately $SO(M)_1$ Kac-Moody algebra~\cite{Tsvelik1}. At $E_J=0$, the effective screening channel operator is the sum $\eta_{i}(0)\eta_{j}(0)+\rho_{i}(0)\rho_{j}(0)$, hence this Hamiltonian is equivalent to $SO(M)_2$ topological Kondo Hamiltonian \cite{Tsvelik1}. However, Eq.~(\ref{HEFF1}) shows that the Josephson coupling $E_J$ results in  channel anisotropy $\Delta J \equiv J_{\rho}-J_{\eta}=\frac{3E_J t^2}{E_c^2}$, breaking the $SO(M)_2$ symmetry down to $SO(M)_1\times SO(M)_1$. For brevity, we shall refer to the topological Kondo phase as the $SO(M)_2$ phase, and to the low energy phase stabilized by the Josephson coupling as $SO(M)_1$  phase. We will also refer to the later as Andreev NFL phase, due to its conductance properties, see below. A related destabilization of the  $SO(M)_2$ fixed point was recently reported for $M=3$ (corresponding to a crossover from 4-channel to 2CK states) in a related spin chain context in Ref.~\onlinecite{giulianoa}

\begin{figure}[pt]
	\centering
	\includegraphics[scale=0.38]{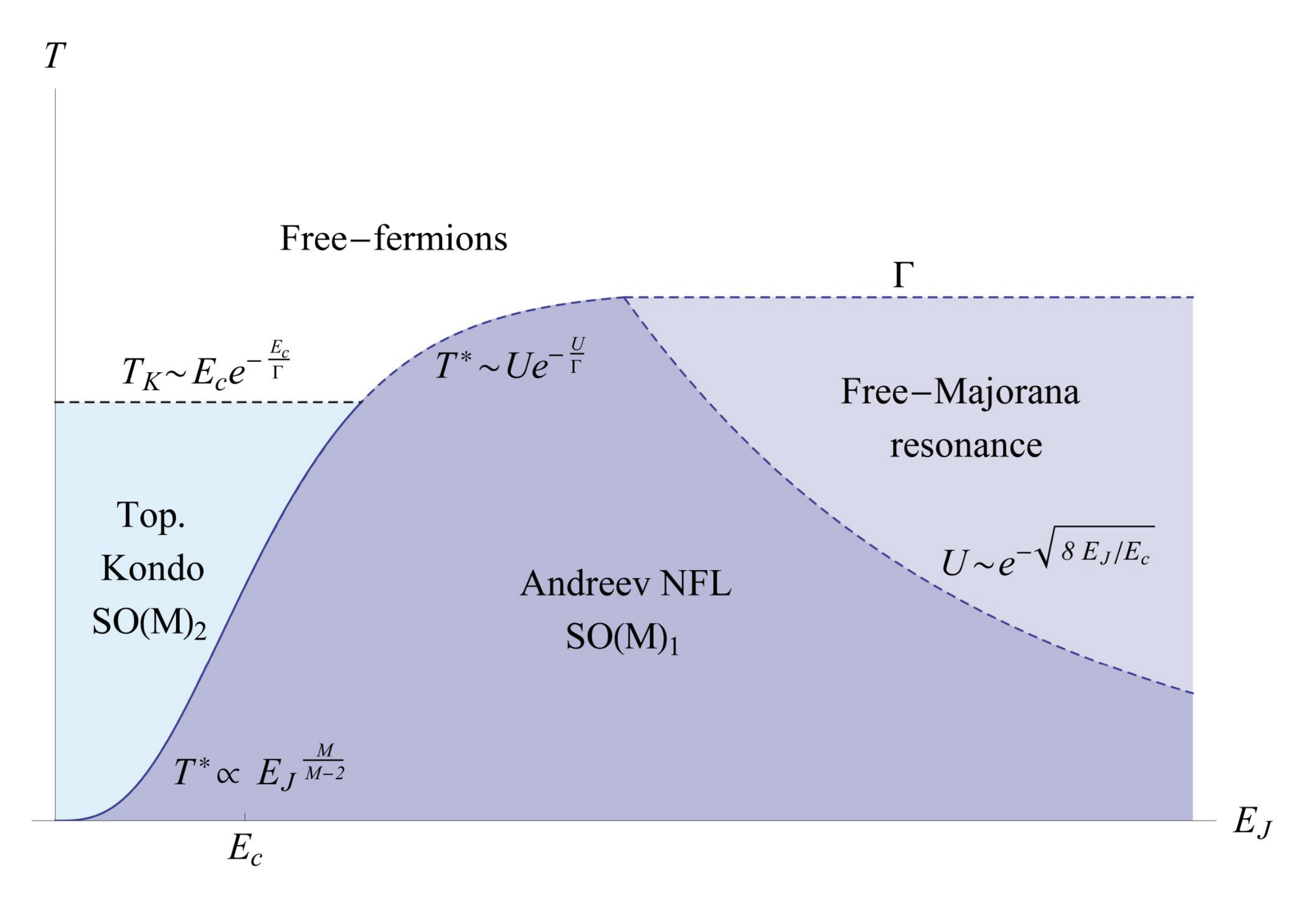}
		\caption{Schematic phase diagram away from a charge degeneracy point. Different sequences of crossovers occur upon decreasing temperature, depending on $E_J/E_c$. In the charging dominated regime, one obtains first a crossover from free fermion behavior, where  the leads are decoupled from the island, to the topological Kondo state, on energy scale $T_K$. This crossover is followed by another one to an Andreev type NFL state, described by $SO(M)_1$, below energy scale $T^*$ given in Eq.~(\ref{Ts}). In the Josephson dominated regime where still $U > \Gamma$, there is a direct crossover from free fermions to the new $SO(M)_1$ phase. When the parity interaction, $U$, is smaller than the tunnel width $\Gamma$, there is first a crossover from free fermions to a non-interacting Majorana resonant state, followed by a second crossover to the $SO(M)_1$ state below $U$. Modification of the phase diagram at a charge degeneracy point is discussed in the text.}
\label{phasediagram}
\end{figure}

As is well known, multichannel Kondo effects are destabilized by channel anisotropy; however, while in topological Kondo setups lead-anisotropy remarkably does not yield any channel anisotropy, we see that the Josephson coupling does lead to channel anisotropy at the topological Kondo fixed point, which is hence unstable. We may start drawing the charging dominated side of the phase diagram of the device, see Fig.~\ref{phasediagram}.

The effect of channel anisotropy on the Kondo fixed point can be extracted by identifying
%. The properties of the fixed point are demonstrated from its operator content, which can be extracted from boundary conformal field theory (CFT). Specifically, we look at $M=3$ at $E_J=0$ where the Hamiltonian at Eq. \ref{HEFF1} coincides with the $SO(4)_2\sim SU(2)_4$ one of B{\'e}ri and Cooper. A convenient boundary CFT formulation then proceeds via $SU(2)_4\times U(1)$ spin-charge decomposition. Reading the operator content of the $SU(2)_4\times U(1)$ CFT, we find
a relevant operator with scaling dimension $\Delta=2/M$~\cite{remark}, which corresponds to tunneling of charge $2e$.
 %, whose quantum numbers $j=0$ and $Q=2$ in the spin and charge sectors respectively.
 While in a charge conserving system this operator is disregarded, the presence of the Josephson coupling indeed allows it. Given that this operator involves degrees of freedom from the leads, we expect its dimensionless coupling constant to be $g_0 \sim \frac{\Gamma E_J}{E_c^2}$ to leading order in $E_J$ and in $\Gamma$. Consequently, The system is driven towards a new fixed point, no matter how initially small $E_J$ is. Using standard renormalization group (RG) analysis~\cite{Pustilnik04}, the crossover to strong coupling takes place at scale
 \begin{align}
 \label{Ts}
 T^*=D_0 g_0^{\frac{M}{M-2}},
  \end{align}
  where  $D_0$ is the initial electron bandwidth. This energy scale $T^*$ which vanishes as $E_J\rightarrow 0 $ is contracted in Fig.~\ref{phasediagram} with the finite Kondo temperature $T_K\sim E_c e^{-\frac{E_c}{\Gamma}}$ signifying the crossover to the low temperature $SO(M)_2$ topological Kondo phase. As our analysis shows, a second crossover necessarily occurs at lower temperature $T<T^*$ into a $SO(M)_1$ phase whose properties will be discussed In Sec. \ref{Low energy conductance}.
\\

\subsection{Parity interaction} In the previous subsection we essentially integrated out the bulk SC and generated effective anomalous couplings between the leads, see Eq.~(\ref{HEFF0}). We now follow the same ideology, this time keeping an internal degree of freedom of the island which can be identified with its parity. Keeping this degree of freedom explicitly is crucial either in the Josephson dominated regime $E_J \gg E_c$ or near resonance $n_g \approx N_0+ \frac{1}{2}$.

\begin{figure}[pt]
	\centering
	\includegraphics[scale=0.26]{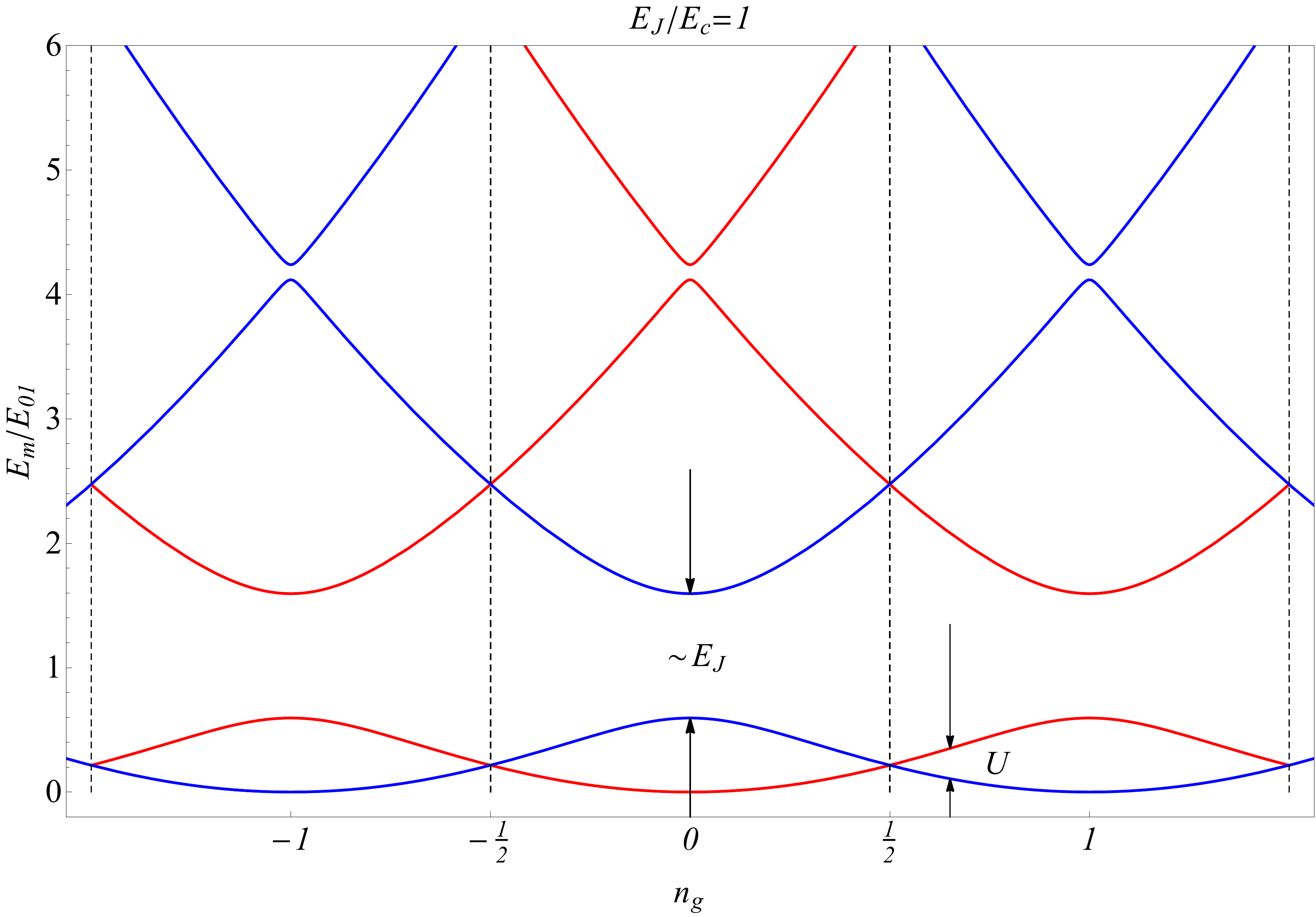}
		\caption{Energy levels of  $H_{\rm{c}}+H_{\rm{J}}$ for $\mathcal{N}$ even (red) or $\mathcal{N}$ odd (blue). There are two lowest energy states of opposite parity, whose energy splitting $U$ is controlled by the gate voltage $n_g$, see Eqs. (\ref{HP}) and (\ref{HUEC}). Here $E_{01}=\rm{min}_{n_g}(E^+_1-E^+_0)$. }
\label{fig2}
\end{figure}

For the moment, let us  focus only on the Hamiltonian of the island together with the Josephson coupling to the bulk SC, $H_{\rm{c}}+H_{\rm{J}}$. As already noted, in the absence of Majorana fermions, the island is a conventional superconductor which is allowed to have only an even number of electrons $\mathcal{N}={\rm{even}}$. The Hamiltonian $H_{\rm{c}}+H_{\rm{J}}$ in this case has a discrete set of eigenstates labeled $m=0,1,2,\ldots$ (which can be expressed in terms of Mathieu functions, see Ref.~\onlinecite{Koch}), depending on the values of $E_J$, $E_c$, and $n_g$, see Fig.
~\ref{fig2}, where these levels are shown as red lines. However, the presence of Majorana modes in the island changes this picture. First, there is an additional set of eigenstates associated with an odd number of electrons, $\mathcal{N}={\rm{odd}}$, see blue lines in Fig. \ref{fig2}. Furthermore, the Majorana modes give rise to degeneracy  $2^{N/2-1}$ of each parity sector (odd or even), where $N \ge M$ is the number of Majorana fermions. We label the energy eigenstates of $H_{\rm{c}}+H_{\rm{J}}$ as $E^p_{m}$, where $m=0,1,\ldots$ and $p=+$ or $-$ for the even and odd sectors of parity respectively.

\begin{figure}[pt]
	\centering
	\includegraphics[scale=0.5]{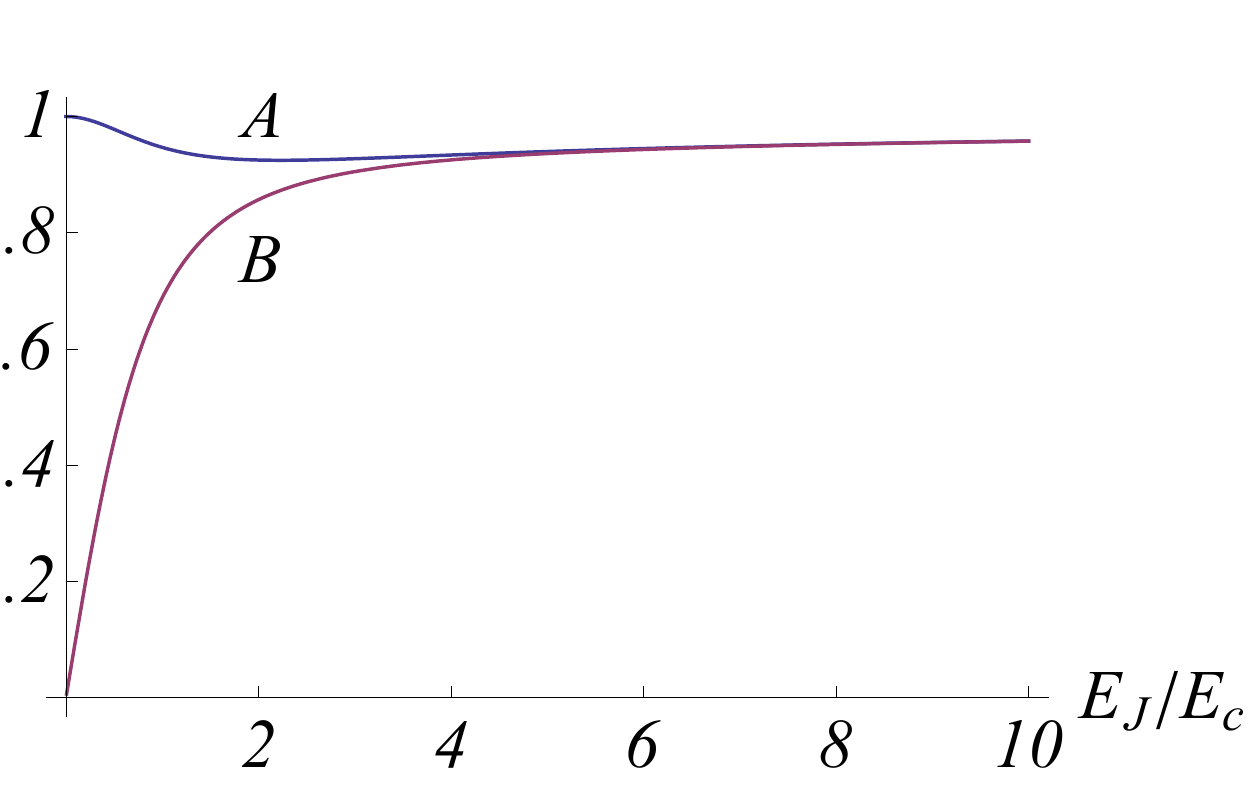}
		\caption{Matrix elements $A$ and $B$ appearing in the effective model Eq.~(\ref{HU}). They are calculated using Mathieu functions (see Ref.~\onlinecite{Koch}) in the on-resonant case $n_g=1/2+{\rm{int}}$. In the charging dominated regime this matches the coefficients in Eq.~(\ref{HUEC}) while in the Josephson dominated regime $A,B \to 1$ with exponentially small difference.  }
\label{fig_AB}
\end{figure}

We shall consider temperatures $T \ll \max\{E_c,E_J \}$. This implies temperatures smaller compared to the excitations gaps of $H_c+H_J$ inside each parity sector, but it allows for two possibilities: (i) low energy subspace with unique parity. As seen in Fig.~\ref{fig2}, this emerges in the charge dominated regime $E_c \gg E_J$ and away from resonance. This situation was considered in the previous subsection. (ii) Quasi-degenerated low energy states with parity $p=\pm$, realized in the Josephson dominated regime $E_J \gg E_c$, or near a resonance $n_g \approx N_0+\frac{1}{2}$. In the latter case, the Hamiltonian can be projected down to a subspace of two lowest eigenstates $|+\rangle$ ,$|-\rangle$ of $H_{\rm{c}}+H_{\rm{J}}$, with eigenvalues $E^+_{0}$, $E^-_{0}$ respectively. Denoting them by a pseudo-spin $\sigma^z |\pm\rangle= \pm |\pm\rangle$ the operator $\mathcal{P}=|+\rangle \langle +|+|-\rangle \langle -|$ projects the system to the manifold of these two states. In general, there is a finite energy difference between these states $U=E^+_{0}-E^-_{0}$, see Fig.~\ref{phasediagram}, which is referred as the $\it{parity}$ interaction. Thus, the projected Hamiltonian takes the form
\begin{align}
\label{HU}
\mathcal{P}H\mathcal{P}= \sum_{j=1}^M \left[t_j\psi_j^\dagger(0)\gamma_j(A\sigma^-+B\sigma^+)+ \rm{h.c}\right]-\frac{U}{2}\sigma^z,
\end{align}
where the matrix elements $A=\langle -|e^{-i\phi/2}|+\rangle$ and $B=\langle +|e^{-i\phi/2}|-\rangle$, as well as the parity interaction $U$, are determined by $E_c$, $E_J$, and $n_g$. (Here we return to general $t_j$ which are not necessarily isotropic). In the following, we calculate them explicitly for the various regimes. In fact, these matrix elements can be evaluated using Mathieu functions as described in Ref.~\onlinecite{Koch}, see Fig.~\ref{fig_AB}. In the case of $B=0$, corresponding to the case $E_J=0$, this model was considered in Ref.~\onlinecite{michaeli}.

 We begin discussing the Josephson dominated regime, in which, as shown by Ref.~\onlinecite{heck}, parity interaction emerges as an exponentially suppressed tunneling  of the phase field. Generally, the two terms $H_{c}$ and $H_J$ compete, as the first tends to fix the number of Cooper pairs, while the second favors charge fluctuations. When $E_J$ is the largest energy scale, the superconducting phase $\phi$ tends to be locked in one of the minima of the cosine term in Eq.~(\ref{HJ}), and thus effectively behaves as a particle in harmonic potential. However, tunneling events between different minima (instantons) where $\phi \rightarrow \phi + 2\pi$ lead to the effective low energy parity interaction~\cite{heck}
\begin{align}
U\sim (E_c E_J^3)^{1/4}e^{-\sqrt{8E_J/E_c}}\cos(\pi n_g).
\label{HP}
\end{align}
 Note that $U=0$ when $n_g$ reaches a degeneracy point. As a single electron tunnels from one of the leads into the island, the parity flips, as realized by the operator $\sigma^x$. One can show that in Eq.~(\ref{HU}) $A=B=1$ up to exponentially small corrections, such that the low energy Hamiltonian yields
\begin{align}
\label{HU2}
\mathcal{P}H\mathcal{P}= \sum_{j=1}^M \left(t_j\psi_j^\dagger(0)\gamma_j \sigma^x+ \rm{h.c}\right)-\frac{U}{2}\sigma^z.
\end{align}
Assuming for simplicity real $t_j$, this becomes
\begin{align}
\label{HU3}
\mathcal{P}H\mathcal{P}=\sum_{j=1}^M \sqrt{2} i t_j  \rho_j(0) \gamma_j - \frac{U}{2} \sigma^z.
\end{align}
In this case, if furthermore $U=0$, then $\sigma_x$ commutes with the Hamiltonian and the pseudo-spin subspace can be eliminated. One major drawback of the Josephson dominated limit is that the parity energy $U$ is exponentially small in $E_J/E_c$. However, this is not necessarily the case in the charge dominated regime as we now discuss.

Consider the charge dominated regime at two different regimes of gate voltage. First, away from a resonance $n_g\approx N_0-\frac{1}{2}$, the Hamiltonian can be projected to its two lowest states manifold $E_0^+$ and $E_0^-$. We assume without the loss of generality that $N_0$ is even. In order to obtain the effective Hamiltonian we calculate the ground states of the two parity sectors $|+\rangle$ and $|-\rangle$ to first order in  $\frac{E_J}{E_c}$,
\begin{align}
|+\rangle & \approx|N_0\rangle+\frac{E_J}{2E_c}|N_0-2\rangle+\frac{E_J}{6E_c}|N_0+2\rangle, \nonumber \\
|-\rangle &\approx|N_0-1\rangle+\frac{E_J}{2E_c}|N_0+1\rangle+\frac{E_J}{6E_c}|N_0-3\rangle.
\end{align}
Using Eq.~(\ref{HU}), we obtain $A\approx 1$, $B\approx \frac{E_J}{E_c}$. The projected Hamiltonian then takes the form,
\begin{align}
\label{HUEC}
\mathcal{P}H\mathcal{P} &= \sum_{j=1}^M \left[t_j\psi_j^\dagger(0)\gamma_j(\sigma^- +\frac{E_J}{E_c}\sigma^+)+ \rm{h.c}\right]-\frac{U}{2}\sigma^z, \nonumber \\
U&=2E_c(n_g-N_0+\frac{1}{2}).
\end{align}
As anticipated, in this case the parity interaction is of order $E_c$.

The situation is more complex in the vicinity of off-resonance point $n_g\approx 0$, where additional excited states are very close to the two-state manifold. For $E_J \ll E_c $ the energy gap $E_{01}^{-}\equiv E_{1}^{-}-E_{0}^{-}\approx 2E_{J}$  is small compared to the gap $ U\approx E_{c}$. This enables transitions from the ground state $E_{0}^{+}$ to $E_{1}^{-}$ , which is very close to $E_{0}^{-}$. As a result, the picture of the pseudo-spin two state manifold collapses. However, for $\frac{E_{J}}{E_{c}}$ of order unity,  $E_{01}^{-}$ already exceeds $U$, see Fig.~\ref{fig3}. Consequently, the pseudo-spin picture holds in this regime. Calculating the coefficients, we find $A=B=\frac{1}{\sqrt{2}}$ such that the projected Hamiltonian has the exactly the same form as Eq.~(\ref{HU3}), where here the parity energy is of order of $E_c$.

For temperatures lower than $U$ the parity of the Hamiltonian Eq.~(\ref{HU2}) is fixed. In order to obtain an effective Hamiltonian in this regime one needs to consider processes in which, after a single electron tunnels from one of the leads into the box, a second electron has to either tunnel $\it{in}$ or out of it. Performing Schrieffer-Wolff transformation starting from the Hamiltonian Eq.~(\ref{HUEC}), we obtain an effective Kondo Hamiltonian exactly as in Eq.~(\ref{HEFF1}),
\begin{align}
H_{\rm{eff}}= \sum_{i\neq j} J_{ij}\rho_i(0)\rho_j(0)\gamma_i\gamma_j,
 \label{HEFF2}
\end{align}
where the exchange coupling are given by $J_{ij}=\frac{2t_i t_j}{U}$, and the Majorana modes $\eta_j(x)$ of the leads decouple from $H_{\rm{eff}}$. This Hamiltonian coincides with Eq.~(\ref{HEFF1}) in the infinite anisotropy limit when $J_\eta=0$; following the above RG analysis, it is obtained as an effective Hamiltonian starting from Eq.~(\ref{HEFF1}) below the energy scale $T^*$, where one of the two $SO(M)_1$ channels decouples.

We now return to the phase diagram, Fig. \ref{phasediagram}, consider the regime $E_J \gtrsim E_c$, and connect it with the small $E_J$ regime discussed earlier. One can associate a Kondo scale $Ue^{-\frac{\Gamma}{U}}$ at which the coupling Eq.~(\ref{HEFF2}) flows to strong coupling. We identify this crossover with the same scale $T^*$ discussed already at small $E_J$ signaling the flow from from the $SO$ into the $SO(M)_1$ phases. Since $U < E_c$, the scale $T^* \sim e^{-\frac{\Gamma}{U}}$ may exceed the Kondo scale for $\Gamma \ll U$.

On the contrary, as the temperature raises above $U$, the effect of the parity interaction becomes unnoticeable such that the system is effectively in the on-resonance regime. At $U=0$ the system consists of $M$ non-interacting Majorana fermions. The coupling $\Gamma$ of each Majorana to a corresponding lead, gives rise to a Majorana resonant state, which forms for temperatures $T \ll \Gamma$, as denoted in Fig.~\ref{phasediagram}.

We briefly speculate on the modification of the phase diagram when the gate voltage is tuned to a charge degeneracy point. In this case the topological Kondo state emerges at scale $\Gamma$ (which exceeds $T_K$). Since the on- and off-resonant Kondo states are described by the same fixed point~\cite{michaeli,mora}, we conclude that the same instability of the topological Kondo state occurs at scale $T^*$ given by Eq.~(\ref{Ts}) for small $E_J/ E_c$.  For large $E_J / E_c$, even though $U=0$, there is a similar crossover between the phase of $M$-decoupled free Majorana resonances, to the $SO(M)_1$ phase, on an exponentially small scale. This energy scale is proportional to the difference $A^2-B^2$ in Eq.~(\ref{HU}), and is identified using the mapping to quantum Brownian motion in a periodic potential below.

\section{Low energy conductance} \label{Low energy conductance} We now probe the low energy properties of the system, including its low temperature conductance, sensitivity to lead asymmetry and to the gate voltage. We will begin this section by a brief review of mapping, which we will then apply to obtain the different fixed points of our system, and then to find their conductance properties.

\subsection{Preliminaries} We briefly review the method by Yi and Kane~\cite{YiKane}, mapping our problem to quantum Brownian motion (QBM) of a particle in a periodic potential.

As a first step towards strong coupling analysis, we bosonize the fermionic fields of the leads. The tunneling part $H_{\rm{T}}$ consists of bi-linears of Majorana and fermionic operators $\psi_j^\dagger\gamma_j$ (or $\gamma_j\psi_j$) which we bosonize
\begin{align}
\psi_j^\dagger(x)\gamma_j \sim e^{i\varphi_j(x)},
\end{align}
where $j=1,2,\ldots,M$ (we set the lattice constant to unity). This bosonization procedure is completely equivalent to combining the Majorana oprators $\gamma_i$ with the fermionic Klein factors of each lead $\xi_i$ \cite{AltlandEgger, Beri}. Since all of these bi-linears commute with the Hamiltonian they can be treated as a c-number that can be absorbed into the tunneling amplitude. In terms of the bosonic fields, the imaginary time action of the leads has the form
\begin{align}
S_{\rm{leads}} =\frac{1}{4 \pi}\sum_{j=1}^{M} \int_{-\infty}^\infty dx \int d \tau  \partial_x \varphi_j(v_F\partial_x \varphi_j - i \partial_\tau \varphi_j).
\end{align}
By integrating out all the degrees of freedom away from $x=0$, this action becomes
\begin{align}
S_{\rm{0}}= \frac{1}{(2\pi)^2}\int d\omega|\omega|   |\vec{\varphi}(\omega)|^2.
\end{align}

The single-electron tunneling is described by
\begin{align}
S_{\rm{T}}= \sum_{j=1}^M t_j \int_{-\infty}^{\infty}d\tau  e^{i\varphi_j(0,\tau)}e^{-i \phi/2} +\rm{h.c}.
\end{align}
 At this point, we follow Yi and Kane \cite{YiKane} and identify $\vec{\varphi}=(\varphi_1,\varphi_2,\ldots,\varphi_M)|_{x=0}$ with the momentum of a particle in a strong periodic potential. In this language, $S_T$ is a hopping term which generates tunneling events between potential minima, while $S_0$ describes an Ohmic coupling of the particle to a dissipative bath. The number of electrons in each lead $(n_1,\ldots,n_M)$ corresponds to the position of the particle in $M$-dimensional space.

Starting at weak lead coupling, the particle is located at one of the potential's minima at $n_j=\rm{integer}$, and is able to hop to adjacent minima separated by a vector $\vec{R}_0$ via $S_{\rm{T}}$. Thus, the allowed charge states, corresponding to the potential minima in the QBM space, form a Bravais lattice. The tunneling Hamiltonian can be expressed then as
\begin{align}
\label{ST}
S_{\rm{T}}= \sum_{j=1}^M t_j  e^{i\sqrt{2}\vec{\varphi}\cdot \vec{R}^{(j)}_0}e^{-i \phi/2} +\rm{h.c}.
\end{align}
 The vectors $\vec{R}^{(j)}_0$ have $M$ components, where only the $j$-th of them is non-vanishing and given by $\frac{1}{\sqrt{2}}$. Following Refs.~\cite{YiKane,michaeli}, our convention is such that the argument of the above exponent is $\sqrt{2}\vec{\varphi}\cdot \vec{R}^{(j)}_0=\varphi_j$, and its scaling dimension is $|R_0|^2=\frac{1}{2}$.

Next, we consider the strong coupling limit of the QBM action, where the bosonic phases $\varphi_i$ are pinned. This corresponds to vanishing of the periodic potential, leading to QBM in free space. The stability of this strong coupling fixed point can be analyzed by examining the effect of a weak periodic potential which has the same periodicity as the original Bravias lattice. Using Fourier decomposition, it is described by $U(r)=\sum_{\vec{G}}v_{\vec{G}}e^{i\vec{G}\cdot r}$, where $\{\vec{G}\}$ is reciprocal lattice vector satisfying $\vec{G}\cdot \vec{R}=\rm{integer}$ for any Bravias vector $\vec{R}$. %Thus, this potential generates tunneling events between minima of Eq.~(\ref{HQMB1})??. 
The scaling dimension of $v_{\vec{G}}$ is given by $|{\vec{G}}^2|$, where we denote the shortest reciprocal lattice vectors by ${\vec{G}_0}$. The length of ${\vec{G}_0}$ determines the leading temperature corrections to physical quantities, e.g., the conductance, as we discuss below.

In conclusion of this part, our model gives rise to low energy fixed points whose low energy properties will be described using the QBM mapping. Different fixed points correspond to different lattices, yielding different leading irrelevant operators. These various options are described in this subsection and summarized in Table I.

\subsection{Fixed points and leading irrelevant operator}
Now we would like to explore the low temperature properties of the various regimes presented in the previous sections. First, we consider the charge dominated regime, where virtual charge transitions give rise to an effective Kondo Hamiltonian, see Eqs. (\ref{HEFF0},\ref{HEFF1}). In QBM language, this Hamiltonian reads
\begin{align}
\label{HQMB1}
H_{\rm{eff}} = \sum_{i\neq j}^M  (J_{\parallel} e^{i\sqrt{2}\vec{\varphi}\cdot \vec{R}^{(ij)}_\parallel}+J_{\perp}e^{i\sqrt{2}\vec{\varphi}\cdot \vec{R}^{(ij)}_\perp})+\rm{h.c},
\end{align}
where $J_{\parallel}=\frac{t^2}{E_c}$, $J_{\perp}=\frac{3t^2E_J}{2E_c^2}$ ($E_c\gg E_J$), and $\vec{R}_{\perp,\parallel}^{(ij)}$ are defined such that $\sqrt{2}\vec{\varphi}\cdot \vec{R}^{(ij)}_\parallel=\varphi_i-\varphi_j$ and $\sqrt{2}\vec{\varphi}\cdot \vec{R}^{(ij)}_\perp=\varphi_i+\varphi_j$.
%\begin{align}
%&J_{\parallel}=\frac{t^2}{E_c}, ~~~ J_{\perp}=\frac{3t^2E_J}{2E_c^2}~~~\rm{for}~~E_c\gg E_J  \\
%&J_{\parallel}=J_{\perp}=\frac{t^2}{U}~~~\rm{for}~\rm{parity}~\rm{interaction}
%\end{align}

These $M$-dimensional vectors $\vec{R}^{(ij)}_\parallel$ and $\vec{R}^{(ij)}_\perp$ in the above equation, correspond to two distinct types of particle hopping in the periodic potential. Specifically, $\vec{R}^{(ij)}_\parallel$ corresponds to charge conserving particle hopping $\psi^\dagger_i\psi_j$, such that  its components sum to $0$, e.g., $\vec{R}^{(12)}_\parallel=\frac{1}{\sqrt{2}}(1,-1,0,\ldots)$. On the other hand, $\vec{R}^{(ij)}_\perp$ corresponds to two electrons tunneling $\psi^\dagger_i\psi^\dagger_j$ described by $i$-th and $j$-th coordinates with the same sign, e.g. $\vec{R}^{(12)}_\perp=\frac{1}{\sqrt{2}}(1,1,0,\ldots)$. Note that the vectors $\vec{R}^{(ij)}_\parallel$ are linearly dependent on  $\vec{R}^{(ij)}_\perp$.

\begin{figure}[pt]
	\centering
	\includegraphics[scale=0.7]{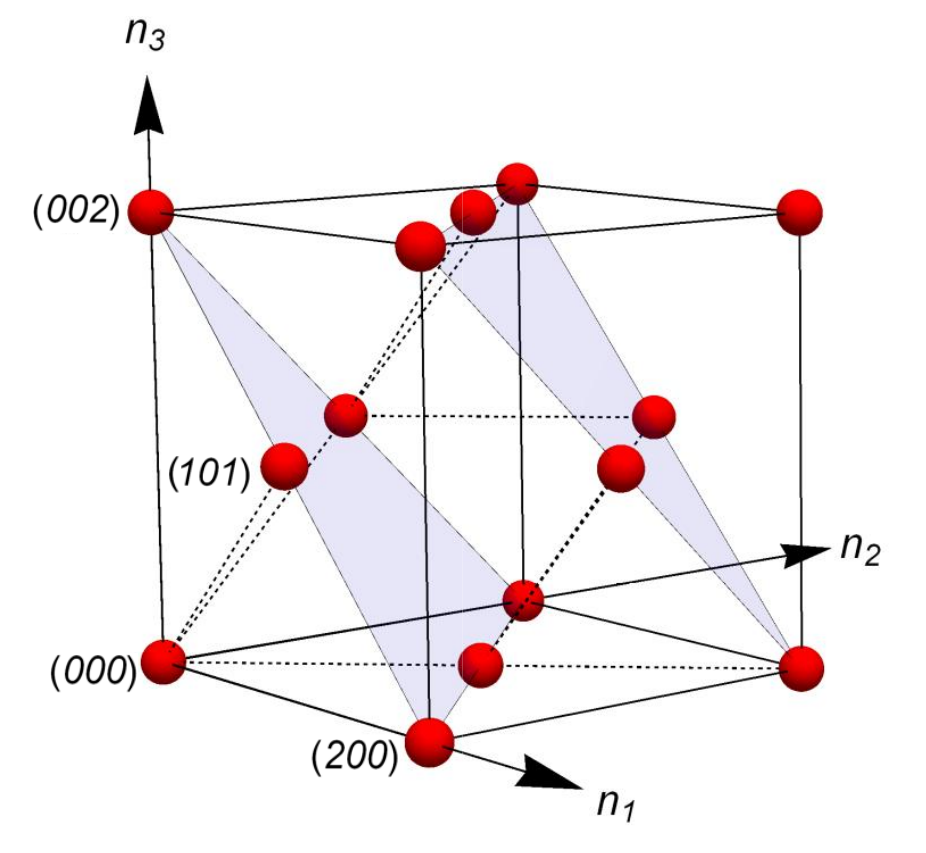}
		\caption{Bravais lattice formed for $M=3$ leads in the Coulomb valley. In the absence of Josephson coupling the allowed charge configurations of the leads form triangular lattices, shown as gray planes, within which the charge is fixed. Turning on $E_J$ allows inter-planar charge transitions, shown as dashed lines. Since parity is conserved the basis vectors are of the form $\sim\left\{1,1,0\right\}$ and thus form an FCC lattice.   }
\label{fig3}
\end{figure}

At $E_J=0$, $J_{\perp}$ vanishes and as a result the motion of the Brownian particle is restricted to an $M-1$ dimensional space, in which the overall charge of the leads $e \sum_j n_j$ is fixed. As already noted, the allowed charge states in this space form a Bravais lattice; specifically, for $M=3$, the particle's motion is restricted to a two dimensional triangular lattice, see gray planes in Fig.~\ref{fig3}. In this case there is no hopping between these planes. The analysis of this system using the QBM mapping was performed in Ref.~\onlinecite{Beri}, leading to a triangular reciprocal lattice, with the resulting leading irrelevant operator $\Delta_M^{(E_J=0)}=2(M-1)/M$.

Crucially, at any $E_J\neq 0 $ tunneling events of two electrons into (or out of) the island generate a finite probability of particle hopping between two parallel planes $\sum_j n_j \rightarrow\sum_j n_j\pm 2$, see dashed lines in Fig.~\ref{fig3}. For $M=3$ this leads to $\it{three}$ dimensional QBM on an FCC lattice, whose basis vectors are $\frac{1}{\sqrt{2}}\left\{1,1,0\right\}$. One should notice that in the case $E_J \ll E_C$, hopping between generalized (1,1,1) planes, $J_{\perp}$, is weaker than the hopping within the planes, $J_{\parallel}$. On the other hand, when the system is dominated by the parity interaction,  $J_{\parallel} \approx J_{\perp} = \frac{2t^2}{U}$, see Eq.~(\ref{HEFF2}). In either case, both tunneling amplitudes are marginally relevant and flow to the same strong coupling fixed point which we now analyze.

Dealing with  the strong coupling limit, the form of the Bravais lattice vectors $\vec{R}^{(ij)}_\perp$ gives the following possible reciprocal lattice vectors: (i) $\vec{G}= \frac{1}{\sqrt{2}}\left(1,1,\ldots,1\right)$, corresponding to the diagonal lattice vector of a (hyper) BCC lattice, with length $|{\vec{G}}|=\sqrt{\frac{M}{2}}$, (ii) $\vec{G}=\left\{\sqrt{2},0,\ldots\right\}$ with length $|{\vec{G}}|=\sqrt{2}$. Therefore, the shortest reciprocal lattice vector has length $|{\vec{G_0}}|=\sqrt{\frac{M}{2}}$ for $M=2,3$, and  $|{\vec{G_0}}|=\sqrt{2}$ for $M>3$. This implies that $v_{\vec{G}}$ is irrelevant for all $M>2$ and marginal for $M=2$. Note that $M=3$ corresponds to 2CK state where the well known scaling dimension of the leading irrelevant operator is $\Delta=\frac{3}{2}$. In conclusion, the weak potential $U(r)$ vanishes during the RG flow, resulting in free space QBM.

We now turn to the on-resonance case where $n_g\approx\rm{integer}+1/2$. The charge conserving case, $E_J=0$, was analyzed in Refs.~\cite{michaeli,mora}. Due to the charge degeneracy between states with $N_0$ and $N_0+1$ electrons in the island, the total charge in the leads $\sum_j n_j$ is permitted to fluctuate by $1$. Consequently, the particle is allowed to hop between \emph{two} adjacent lattice planes perpendicular to the direction $\frac{1}{M}(1,1,\ldots,1)$. For $M=3$, the formed lattice is a corrugated honeycomb lattice consisting of two triangular sublattices. Note that for $E_J=0$ in the off-resonance case, the particle hops between sites of the triangular lattice via virtual transitions through the high-energy sublattice. The honeycomb lattice, however, has the same Bravais lattice as each triangular lattice; as result of this the structure of the leading irrelevant operator is the same as in the off-resonance case.

\begin{figure}[pt]
	\centering
	\includegraphics[scale=0.7]{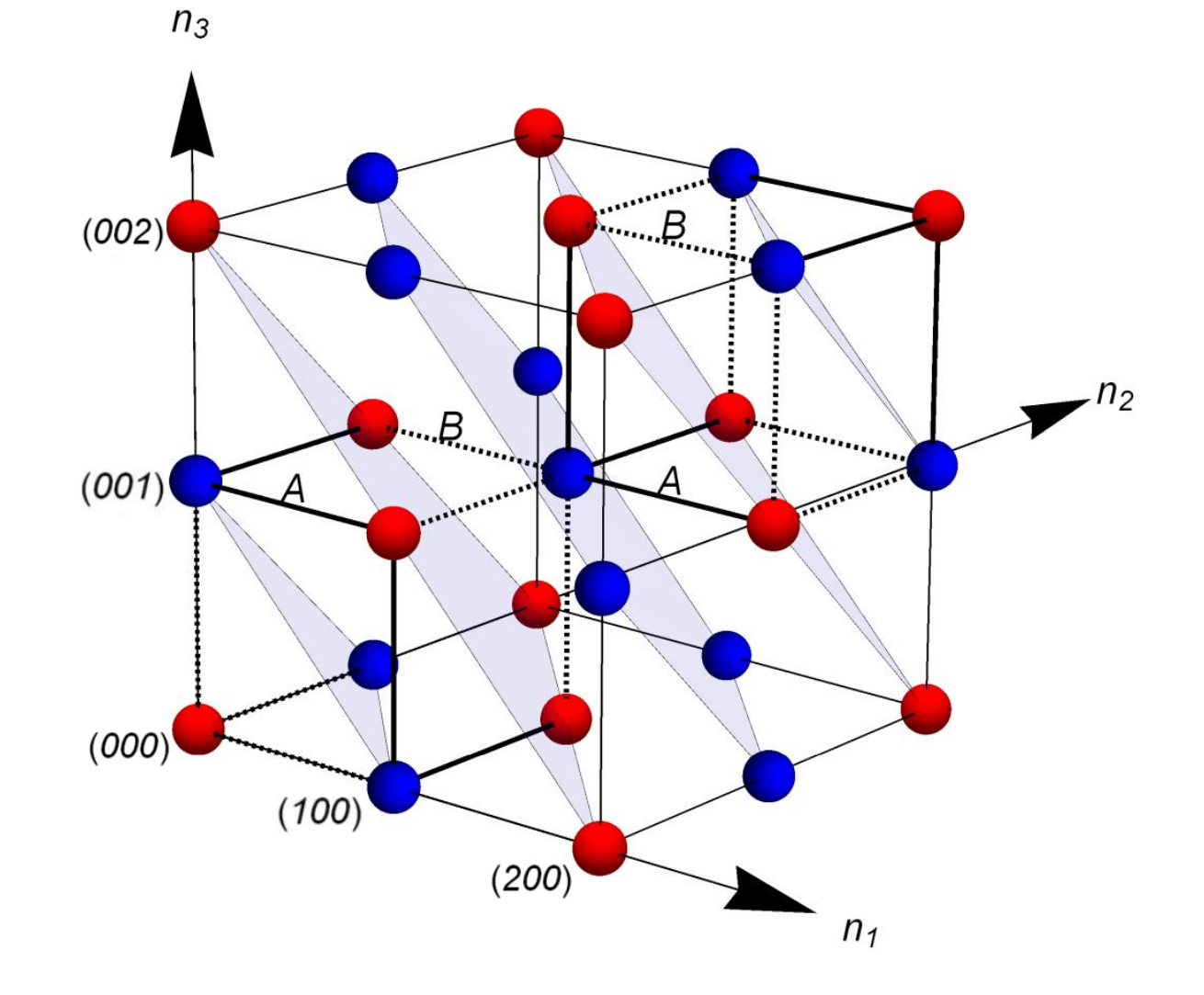}
		\caption{Lattice formed for $M=3$ leads in the on-resonant regime. It can be decomposed into triangular planes characterized by even (red dots) or odd (blue dots) integer value of $n_1+n_2+n_3$. The on-site energy of the two sublattices is the same at $U=0$ (on-resonance). The interplanar charge transitions is now staggered, as in Eq.~(\ref{HU}), see thick ($A$) versus dashed ($B$) lines. The unit cell and Bravais lattice is the same as in Fig.~\ref{fig3}, i.e., FCC. Only in the non-interacting limit $E_c=0$, we have $A=B$ and the lattice becomes simple cubic.}
\label{fig3b}
\end{figure}

At finite $E_J$, the effective tunneling is given by Eq.~(\ref{HU2}). On-resonance $U$ vanishes, allowing fluctuations of the parity. The QBM action then takes the form
\begin{align}
S = S_{\rm{0}}+ \sum_{j=1}^M [t_j  e^{i\sqrt{2}\vec{\varphi}\cdot \vec{R}^{(j)}_0}(A \sigma^-+ B \sigma^+) +\rm{h.c}].
\end{align}
%As for the off-resonance case, the Josephson coupling has a dramatic effect on the structure of the lattice: 
Importantly, at $E_J \neq 0$ tunneling events of $2e$ enable the particle to hop between $(1,1,1)$ planes characterized by any integer $\sum_j n_j$, see Fig.~\ref{fig3b}.
This set of planes may be divided into planes where $\sum_j n_j $ is even or odd, see red and blue lattice sites in Fig.~\ref{fig3b}. While this set of lattice sites forms an (hyper) cubic lattice, for $A \ne B$ there is a staggered structure in the tunneling between planes, see thick versus dashed lines in Fig.~\ref{fig3b}. Consequently, the corresponding Bravais lattice remains FCC as in the off-resonant case.

In this case there is a distinction between the Bravais lattice vectors of FCC, see Table~I, and the shortest lattice vectors appearing in the tunneling Hamiltonian, $\vec{R}^{(j)}_0$, with $|R_0|^2=\frac{1}{2}$, corresponding to (hyper) cubic lattice. Being a relevant perturbation, the tunneling Hamiltonian flows to strong coupling. By analyzing the reciprocal Bravais lattice, i.e., BCC, as a perturbation, we obtain the same scaling dimension of the leading irrelevant operator as in the off-resonance case.

The distinction between the (hyper) cubic and FCC lattices is due to the difference between the tunneling amplitudes $t_A=t\cdot A$ and $t_B=t\cdot B$, where $t_j=t$ is isotropic. This difference, however, vanishes at large $E_J/E_c$.  In fact, in this regime the superconducting phase $\phi$ is localized in the minima of the cosine potential with a typical localization length (in units of $2 \pi$) of $(E_c/E_J)^{1/4}$. The sensitivity of the wave function in the phase representation to boundary conditions, which is measure in the difference between $A$ and $B$, is exponentially suppressed in $2\pi/(E_c/E_J)^{1/4}$, see Fig.~\ref{fig_AB}. Thus, at temperatures higher than an exponentially small energy scale, similar to $U$ in Fig.~\ref{phasediagram}, QBM takes place essentially on a hyper-cubic lattice. This Bravais lattice having the same reciprocal lattice, leads to a leading irrelevant operator of dimension $2$, see Table I.

In all discussed cases, where the electron tunneling flows to strong coupling and hence the periodic potential flows to weak coupling, the effect of lead-anisotropy is seen to be irrelevant.

\begin{table}
    \begin{tabular}{|p{1.7cm}| p{2.1cm} | l |  p{1.7cm} |l|}
    \hline
		system & lattice &$\sqrt{2}\vec{R}^{(0)} $ &  $\vec{G}_0/\sqrt{2} $  & $\Delta_{M} $\\ \hline
    $E_J=0$ \newline $n_g \ne \frac{1}{2}+N_0$ & triangular &$\{1,-1,0\} $&  $\{-\frac{2}{3},\frac{1}{3},\frac{1}{3}\}$  \newline triangular & ~~$\frac{4}{3}$ \\ \hline
    $E_J=0$ \newline $n_g = \frac{1}{2}+N_0$ & honeycomb & $\{1,0,0\}^{*}$ & $\{-\frac{2}{3},\frac{1}{3},\frac{1}{3}\}$  \newline triangular &  ~~$\frac{4}{3}$  \\ \hline
    $E_J \ne 0$ \newline $n_g \ne \frac{1}{2}+N_0$ & FCC & $\{1,1,0\}$ & $\{\frac{1}{2},\frac{1}{2},\frac{1}{2}\} $  \newline BCC & ~~$\frac{3}{2}$  \\ \hline
    $E_J \ne 0$ \newline $n_g = \frac{1}{2}+N_0$ & cubic \newline (FCC Bravais) & $\{1,0,0\}$ &  $\{\frac{1}{2},\frac{1}{2},\frac{1}{2}\}$ \newline BCC & ~~$\frac{3}{2}$ \\  \hline
		$E_J \gg E_c  $ \newline $E_c \to 0 $ & cubic & $\{1,0,0\} $ & $\{1,0,0\} $ \newline cubic & $~~2$\\ \hline
           \end{tabular}
            \caption{Summary of the various lattices and lattice vectors introduced in the QBM description. For clarity, we restrict our attention to $M=3$ leads. At $E_J=0$ lattices are 2 dimensional, while for $E_J \ne 0$ lattices are 3-dimensional. $\vec{R}^{(0)}$ refers to the shortest vector, which determines the scaling dimension of $H_T$ via $|\vec{R}^{(0)} |^2$, which is not necessarily a vector of the Bravais lattice. $\vec{G}_0$ is the shortest reciprocal lattice vector. The (Bravais) reciprocal lattice is denoted below each $\vec{G}_0$ vector. In the lattice vector of the honeycomb lattice $\{1,0,0 \}^*$ motion is restricted to two neighboring $(1,1,1)$ planes.}
        \end{table}

\subsection{Conductance} 
We now discuss the conductance focusing on the new phases stabilized by the Josephson coupling, using the QBM picture applied in the previous sections.
%For equal leads coupling the resulting current at $T=0$ in the presence of Josephson junction yields:
%\begin{align}
%\label{GV}
%I_i = G V_i
%\end{align}
In the strong coupling limit, the QBM takes place in $M$-dimensional free space obtained after the vanishing of the periodic potential during RG flow. Suppose that we apply a voltage $V_1=V$ on a single lead, $i=1$. In the QBM action, the voltage $V_1$ couples to the electron number in lead $1$, $n_1$, as $-eV_1n_1\equiv \mathcal{V}(n_1)$ corresponding to a linear potential in the particle's coordinate $n_1$, i.e., to an electrical field in this direction. This gives rise to a force $F=-\frac{d{\mathcal{V}}(n_1)}{dn_1}$, which, in the presence of dissipation, leads to steady-state velocity $\dot{n}_1$ via $0=\frac{d^2 n_1}{dt^2}=F-\frac{m\dot{n}_1}{\tau}$, where $\tau$ is the mean free time of the Brownian particle. Rather than computing $\dot{n}_1$ we use the same method as \cite{michaeli}, and argue that the steady-state velocity is independent of the dimensionality $M$ due to fact that the free space QBM is spatially isotropic and decoupled along different directions. Thus we conclude that $I_1$ is independent of $M$ (Notice however, that while in the charge conserving situation~\cite{michaeli} the dimensionality equals $M-1$, in our case the Brownian particle can explore all $M$-dimensions). For $M=1$, the current in lead $1$ is given by $I_1=\frac{2e^2}{h}V$ \cite{alicea2}. Therefore, we find $I_1=\frac{2e^2}{h}V$ for all $M$ at zero temperature and for finite $E_J$. In general one can discuss a conductance matrix $G_{ij}$ such that $I_i = \sum_{j}G_{ij} V_j$. At $T=0$, $G_{ij} =\frac{ 2e^2}{h} \delta_{ij}$ for $E_J > 0$.

Low temperature corrections of the conductance are dominated by the leading irrelevant operator of the strong coupling fixed point summarized in Table I. As already noted, this operator follows from weak periodic potential which has (hyper) FCC structure and has a scaling dimension $|\vec{G}_0|^2$. Consequently, we obtain 
\begin{align}
G_{ij} = \frac{2e^2}{h}(\delta_{ij}+A_{ij}T^{2\Delta_M-2}),
\end{align}
where $\Delta_M = \frac{3}{2}$ for $M=3$, or $\Delta_M = 2 $ for $ M>3$, and $A_{ij}$ are non-universal constants depending on the problem's parameters. In the regime of $T > U$, denoted as  $M$-decoupled Majorana resonant states in Fig.~\ref{phasediagram}, we have $\Delta_M=2$ for any $M$.

This universal power law should be contrasted with the result of Eriksson et. al~\cite{Eriksson14}, finding a manifold of fixed points with continuously varying exponents. The latter was achieved (i) in the Josephson dominated regime where the parity interaction is negligible, and (ii) in a special situation where $T_K$  exceeds the tunnel width $\Gamma$ (as opposed to our assumptions, see Fig.~\ref{phasediagram}).

\subsection{Interactions} \label{Interactions}
\begin{figure}[pt]
	\centering
	\includegraphics[scale=0.25]{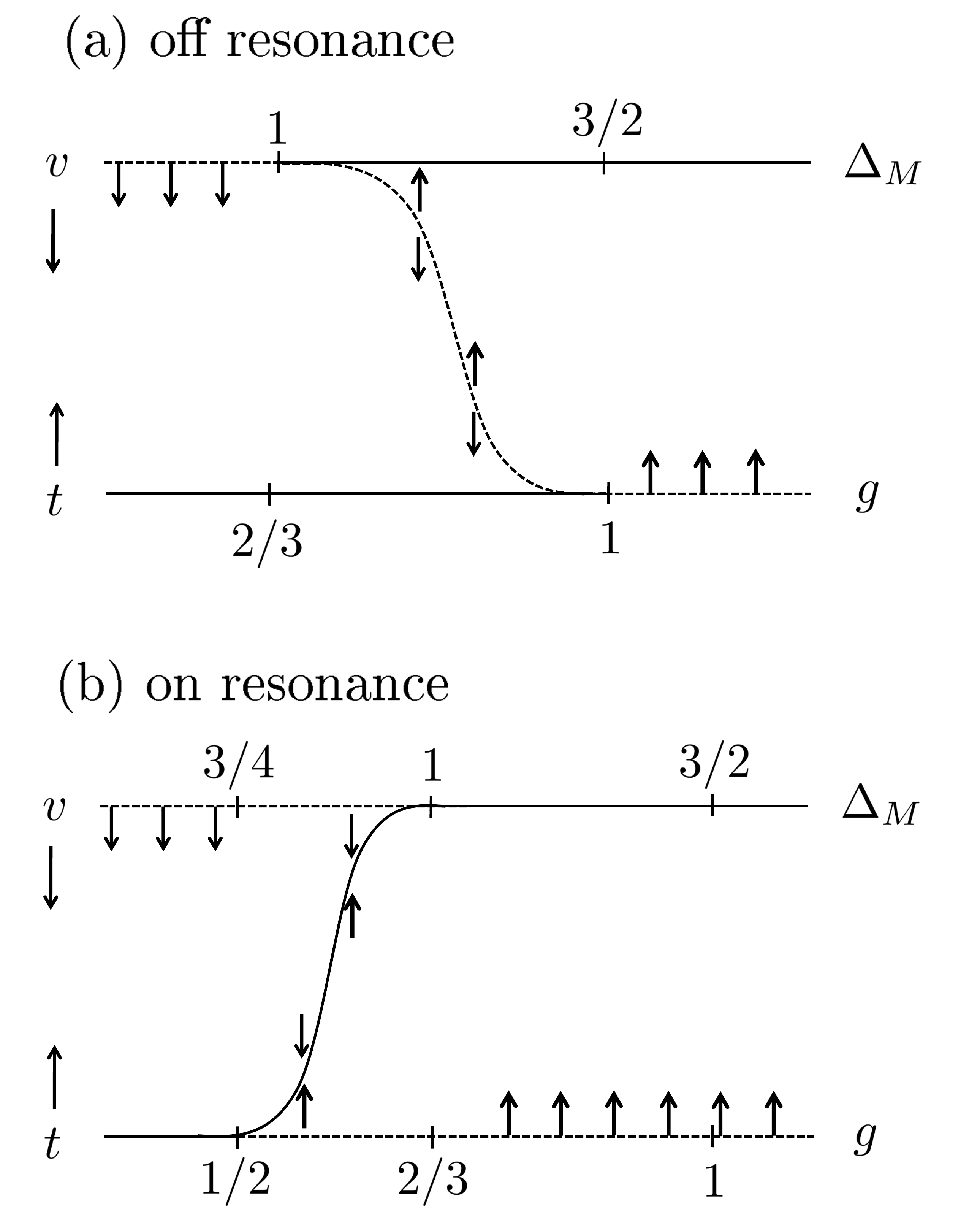}
	\label{fig4}
	\caption{Phase diagram for $M=3$ (a) in the off-resonance case and (b) near a charge degeneracy point, as function of the Luttinger parameter $g$ of the leads. The corresponding lattices in the QBM language are shown in Fig.~\ref{fig3} and Fig.~\ref{fig3b}, respectively. The lower line corresponds to weak leads coupling $t\rightarrow 0$ and the upper line corresponds to the strong coupling regime, where the QBM takes place in a weak periodic potential  $v\rightarrow 0$. Stable (unstable) fixed points are marked by solid (dashed) line. \
		}
\label{fig4}
\end{figure}
Using the QBM formulation, generalization of the previous analysis to interacting leads is straightforward. The interactions are given by the Luttinger parameter $g$. In order to study their effect, we find the change in the length of both the Bravias and the reciprocal lattice vectors, which is given by $|\vec{R}|\rightarrow |\vec{R}|/\sqrt{g} $, $|\vec{G}|\rightarrow \sqrt{g}|\vec{G}| $ \cite{YiKane}.

The phase of the system strongly depends on whether the gate voltage is either on- or off-resonance. For concreteness, we focus on the case $M=3$. First, if the system is off-resonance, we find a line of intermediate unstable fixed points, see Fig. \ref{fig4}. This line emerges since there is a range of $g$ in which both $t$ and $v$ are irrelevant. Explicitly, the Bravais lattice vectors of FCC and BCC, see Table I, give the relation $|R|^2|G|^2=3/2$, implying that at the marginal point of $t$, $|R|^2=1$, $|G|^2=3/2>1$, hence $v$ is irrelevant. 
On the other hand, on-resonance, the scaling dimension of the tunneling operator $t$ is determined by a non-Bravais vector $R_0$. As seen in Table I, $|\vec{R}_0|^2=(|\vec{R}|^2)/2$, hence, in this case the marginal point of $t$, $|R_0|^2=1$, gives $|G|^2=3/4<1$, implying that $v$ is relevant. Thus, in the on-resonant case we obtain an intermediate line of stable fixed points, see Fig.~\ref{fig4}(b).

\section{Summary} \label{Summary}
 To conclude our work, we showed that Josephson coupling gives rise to a substantial change in the physics of Majorana islands. The full phase diagram of the system depending on $\frac{E_J}{E_c}$ and $n_g$ has been obtained, predicting universal values of the conductance at $T=0$ and its power-law low temperature corrections. While the original model including the bulk superconductor is more complicated, the effective model in Eq.~(\ref{HU}) may be used to test our predictions using numerical techniques. With the fast progress in the field, we are optimistic that our predictions will be verified experimentally.

{\it Acknowledgements:} We thank R. Egger, C. Mora, K. Michaeli and L. Fu for helpful and interesting discussions. This work is supported by Israel Science Foundation Grant No.~1243/13, and the Marie Curie CIG Grant No.~618188.

%Summary of the various lattices and lattice vectors obtained depending on charge conservation and degeneracy of charge stares. In the lattice vector of the honeycome lattice $\{1,0,0 \}^*$ motion is restricted to two neibohring $(,1,1,1)$ planes.

\bibliographystyle{apsrev}

\begin{thebibliography}{10}
\bibitem{remark}
	This can be obtained as in Ref.~\onlinecite{michaeli} by decomposing the scaling dimension of the tunneling operator
	$1/2=R^2=R_\perp^2+R_{\parallel}^2$ where $R_\perp^2$ accounts for the charge sector.
\end{thebibliography}

\end{document}